\documentclass[9pt,twocolumn,twoside]{osajnl}
\usepackage[ justification=justified]{caption} % 'format=plain'
\usepackage{multirow}
\usepackage{titlesec}
\graphicspath{{images/}{../images/}}

\journal{optica} % Choose journal (ao,jocn,josaa,josab,ol,optica,pr)

\setboolean{shortarticle}{false}
\usepackage{lineno}
\title{Modulation leakage-free continuous-variable quantum key distribution }

\author[1,*]{Adnan A.E. Hajomer}
\author[1]{Nitin Jain}
\author[1]{ Hossein Mani}
\author[1,2]{Hou-Man Chin}
\author[1]{Ulrik L. Andersen}
\author[1,$\dagger$]{Tobias Gehring}

\affil[1]{Center for Macroscopic Quantum States (bigQ), Department of Physics, Technical University of Denmark, 2800 Kongens Lyngby, Denmark}
\affil[2]{Department of Photonics, Technical University of Denmark, 2800 Kongens Lyngby, Denmark}
\affil[*]{Corresponding authors: * aaeha@dtu.dk, $^\dagger$ tobias.gehring@fysik.dtu.dk}

\begin{abstract}
Distributing cryptographic keys over public channels in a way that can provide information-theoretic security is the holy grail for secure communication. This can be achieved by exploiting quantum mechanical principles in so-called quantum key distribution (QKD). Continuous-variable (CV) QKD based on coherent states, in particular, is an attractive scheme for secure communication since it requires only standard telecommunication technology that can operate at room temperature. However, a recently discovered side-channel created in the process of state preparation leads to a leakage of information about the transmitted quantum state, opening a security loophole for eavesdroppers to compromise the security of the CVQKD system. Here, we present a CVQKD system without this modulation leakage vulnerability. Our implementation is based on a baseband modulation approach, and uses an in-phase and quadrature (IQ) modulator for state preparation and radio frequency heterodyne detection together with carefully designed digital signal processing for state measurement. We consider practical aspects in the implementation and demonstrate the generation of a composable secret key secure against collective attacks. This work is a step towards protecting CVQKD systems against practical imperfections of physical devices and operational limitations without performance degradation.  
\end{abstract}

\setboolean{displaycopyright}{true}

\begin{document}

\maketitle

\section{Introduction}
Quantum key distribution (QKD) allows communicating parties to securely distribute cryptographic keys by harnessing the fundamental properties of quantum mechanics. Theoretically, using QKD together with one-time pad encryption provides information-theoretic secure data transmission that cannot be broken either by current or by future technology~\cite{scarani2009security,pirandola2020advances}. However, any practical realization of QKD is vulnerable to side-channels, resulting mainly from imperfections of physical devices and operational limitations, which may not be taken into account in an idealized security proof. %In discrete variable QKD, for instance, such imperfections open security loopholes for eavesdropper Eve and result in different types of attacks
For instance, in discrete variable QKD, typical attacks exploiting side-channels are Trojan horse attacks \cite{Gisin2006, jain2011device} and detector blinding \cite{lydersen2010hacking,gerhardt2011full}.

In continuous variable (CV) QKD, quantum states are encoded into the two orthogonal---phase and amplitude---quadratures of the electromagnetic light field. Prepare-and-measure protocols, where the sender (Alice) prepares coherent quantum states using quadrature modulation, are  particularly common as the CVQKD system can be built from telecommunication components~\cite{pirandola2020advances,zhang2020long,grosshans2002continuous,weedbrook2004quantum,laudenbach2018continuous,jouguet2013experimental}. The prepared quantum states are  transmitted through an insecure quantum channel, ased to be fully controlled by an eavesdropper (Eve). At the receiver (Bob), the quantum states are measured in a coherent manner, e.g., using homodyne or heterodyne detection \cite{weedbrook2012gaussian,kikuchi2015fundamentals}. Finally, the QKD protocol ends with classical data processing and security analysis, characterizing the information advantage of Alice and Bob over Eve. Similar to its discrete variable counterpart, CVQKD is prone to implementational loopholes. For example, attacks on the local oscillator's intensity and the shot noise calibration have been demonstrated \cite{ma2013local,jouguet2013preventing,qin2016quantum,huang2013quantum}. 

%One approach to cope with side-channels due to imperfections of physical devices, is to make a security proof completely independent of  devices \cite{acin2007device}. This approach is, however, experimentally challenging and results in a very low secret key rate \cite{thearle2018violation,acin2007device}. As a compromise  between the security and  performance of the protocol, the idea of removing  asptions on devices has been realized  partially in measurement-device-independent (MDI) QKD \cite{pirandola2015high,li2014continuous}, where the detector is in full control of Eve, while the state preparation devices are completely trusted. Therefore, MDI QKD is also  subject to side-channels due to imperfections of  quantum states  preparation devices.

 Recently, the impact of side-channel leakage during coherent state preparation in a CVQKD protocol has been theoretically analysed and experimentally demonstrated in a setup that established the coherent state's mode in an optical single sideband (OSSB) using in-phase and quadrature (IQ) modulation~\cite{derkach2017continuous,jain2021modulation}. In the proof-of-concept experiment the authors showed that a side-channel is created due to the finite suppression of the quantum-information-carrying image sideband during OSSB modulation. This side-channel can significantly reduce the secret key rate and can even break the security of the system if neglected in the security proof. Therefore, exploring a new modulation scheme for quantum state preparation which is information-leakage-free, is of great importance to CVQKD implementations. 
 
 Here we report a CVQKD implementation that is capable of removing the modulation leakage vulnerability without considering the practical imperfections of the modulation process in the security proof. We achieve this by implementing an optical baseband modulation scheme using an IQ modulator for quantum state preparation, and radio frequency (RF) heterodyne detection. In contrast to previous implementations \cite{kleis2017continuous,qu2016rf,chin2021machine,brunner2017low,jain2021modulation}, the proposed modulation scheme does not frequency shift the coherent states at a radio frequency but keeps the quantum signal band centered at the optical carrier frequency. Thereby, it removes the side-channel leakage from the suppressed (image) sideband of quantum signals. Simultaneously, our scheme still employs a continuous-wave laser and digital mode shaping and does not require an additional amplitude modulator for pulse carving. We experimentally evaluate the performance of our implementation in terms of the so-called excess noise~\cite{pirandola2020advances, laudenbach2018continuous}: with the help of digital signal processing (DSP) that includes an optimized high pass filter and a machine learning framework for phase carrier recovery~\cite{chin2021machine}, our implementation achieved one of the lowest-ever reported excess noise of $0.72 \times 10^{-3}$ photon number units (PNU). In addition, we also achieved a  composable secure key fraction~\cite{gehring2015sqzcomposableqkd, jain2021practical} of 0.007 bits/symbol, calculated under the asption of collective attacks. 
 
 We note that the proposed baseband modulation is also suitable for other CVQKD systems based on modulated coherent states, including not only point-to-point CVQKD but also measurement-device-independent CVQKD~\cite{li2014continuous}. Therefore, it could become the defacto scheme for future CVQKD implementations.  

\section{Coherent state preparation}
\label{sec:state preparation}

 To prepare coherent states, quadrature modulation is used in CVQKD systems \cite{grosshans2002continuous}. In practice, quadrature modulation can be implemented either by using discrete amplitude and phase modulators or an IQ modulator~\cite{lance2005no,zhang2020long, kleis2017continuous,qu2016rf}. The latter option offers more compact design, cost-effectiveness and potential robustness against Trojan-horse attacks~\cite{jain2014risk, jain2021modulation}. Since in the IQ modulator the electro-optical modulators are fabricated on a single substrate, they have similar electrical-to-optical transfer functions which is an important property for generating coherent states in OSSBs.
 
\begin{figure}[t]
\centering
%\fbox{\includegraphics[width=\linewidth]{IQmodulator.pdf}}
\includegraphics[width=\linewidth]{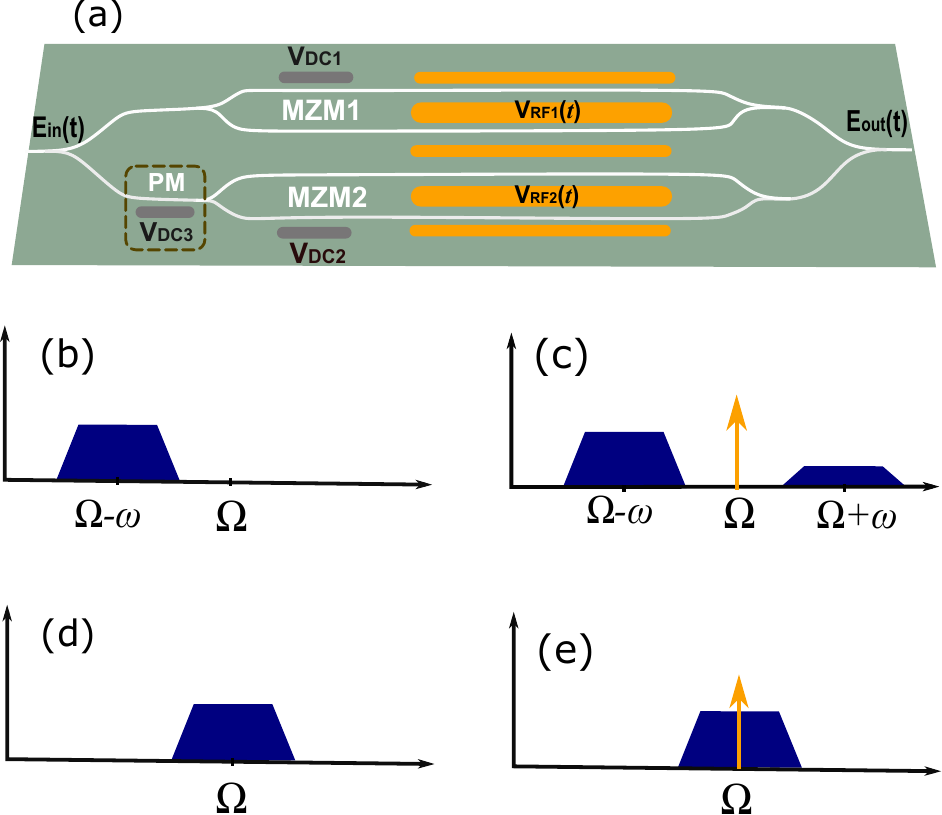}
\caption{IQ modulator internal structure and its operation modes. (a) An IQ modulator consists of two MZMs driven by RF signals $V_\text{RF1}(t)$ and  $V_\text{RF2}(t)$ with $90^{\circ}$ phase difference. For OSSB-CS, the two MZMs operate at a dark fringe, e.g., through $V_\text{DC1} = V_\text{DC2} = V_\pi$, while the PM maintains the relative phase of $90^{\circ}$ using $V_\text{DC3}$. (b)-(e) Output spectra of OSSB and baseband operation modes. (b) Infinite carrier and sideband suppression, and therefore ideal OSSB-CS. (c) OSSB-CS with finite carrier and sideband suppression due to sub-optimal settings of the DC bias and manufacturing imperfections. (d) Ideal baseband modulation with carrier suppression. (e) Practical baseband modulation with finite carrier suppression.}
\label{fig:1}
\end{figure}

 The IQ modulator is, in principle, a Mach–Zehnder interferometer with dual nested Mach-Zehnder modulators (MZMs) and a phase modulator (PM), as shown in Fig.~\ref{fig:1}(a). 
 Given an electric field $E_\text{in}(t)$ at the input of an ideal IQ modulator, the electric field at the output can be expressed as~\cite{Izutsu1981IntegratedIQmod, li2017modulation}, 
\begin{multline}
E_\text{out} (t) = \frac{E_\text{in}(t)}{2}\left[\cos\left(\pi \frac{ V_\text{RF1}(t)+V_\text{DC1}}{2V_\pi}\right)+ \right.\\  \cos\left(\pi \frac{V_\text{RF2}(t)+V_\text{DC2}}{2V_\pi}\right) e^{-j \pi\left(\frac{V_\text{DC3}}{2V_{\pi/2}}\right)} \biggr]\ , 
\label{Eq:1}
\end{multline}
 where $V_\text{RF1}(t)$ and $V_\text{RF2}(t)$ are RF waveforms and $V_{\text{DC}{1,2,3}}$ are direct current (DC) bias voltages applied on the modulators. The latter are usually controlled by an automatic bias controller (ABC). Here, $V_\pi$ is the half-wave voltage, i.e., the voltage at which the optical phase changes by $\pi$. In CVQKD, the modulator is usually operated in the linear regime. Hence, $V_\text{RF1}(t)$ and $V_\text{RF2}(t)$ are typically (at least) an order of magnitude smaller than $V_\pi$ and only the first order sidebands of the electric field get significantly excited.

For CVQKD purposes, OSSB modulation for the generation of coherent states provides practical advantages as it avoids low frequency noise by shifting the quantum states with coherent excitation $\alpha(t) = I(t) + jQ(t)$ away from the optical carrier~\cite{kleis2017continuous,chin2021machine}.
As illustrated in Fig.~\ref{fig:1}(b), in an ideal scenario, the optical carrier at $\Omega$ as well as the (upper) sidebands around $\Omega + \omega$ are completely suppressed, leading to true OSSB modulation with carrier suppression (OSSB-CS). However, in practice, the IQ modulator is capable of providing only a finite level of OSSB-CS as depicted in Fig.~\ref{fig:1}(c). In this practical scenario, the output electric field of equation (\ref{Eq:1}) can be expressed as,
 \begin{align}
    E_\text{out}(t) & \approx \frac{ \mu}{2}\alpha(t)e^{j(\Omega - \omega) t} + \frac{\delta}{2}\alpha^*(t)e^{j(\Omega + \omega) t} + \Delta e^{j\Omega t}\ .
    \label{Eq:4}
 \end{align}
Here, $\mu$ is the modulation index, $\delta < \mu$ is an effective modulation index including the sideband suppression ratio, and $\Delta$ describes the residual electric-field amplitude of the carrier after suppression. A detailed derivation can be found in Ref.~\cite{jain2021modulation}.

The modulation leakage vulnerability gets manifested when Alice and Bob do not account for the finite suppression, i.e.\ $\delta > 0$, while Eve extracts the not-completely-suppressed sideband (centered at $\Omega + \omega$ in Fig.~\ref{fig:1}(d)), using, for instance, an optical filter \cite{jain2021modulation}. Basically, this allows Eve to gain more information about the secret key without alerting Alice and Bob. 

 One way to deal with this side-channel while keeping the advantages of OSSB modulation, is to consider the information leakage from the suppressed sideband  in the security proof. Nevertheless, this approach reduces the range over which a secret key can be obtained, requires additional measurements when characterizing the system and the long-term stability of the achieved suppression has to be guaranteed~\cite{jain2021modulation}. Alternatively, one can completely remove the issue of finite sideband suppression and therefore the modulation leakage vulnerability by considering a modulation scheme in which no suppressed image sideband is generated.
    
 One possible candidate for such a modulation scheme is optical baseband modulation~\cite{agrawal2012fiber}, a technique in which the driving RF waveforms of the IQ modulator are baseband signals, i.e.\ the signals have not been up-converted to a higher radio frequency. This results in the quantum signal band centered at the optical carrier frequency as shown in Fig.~\ref{fig:1}(d). In fact, by driving the IQ modulator such that $V_\text{RF1}(t) \propto \text{Real}(\alpha (t)) = I(t)$ and $V_\text{RF2}(t) \propto \text{Imag}(\alpha(t)) = Q(t)$ and considering the imperfect bias settings of the MZMs, the electric field at the output of the IQ modulator can be expressed as, 
 \begin{align}
    E_\text{out}(t) & = \biggr[\sin \left(\mu_1 I(t)+\phi_1 \right)+j \sin \left(\mu_2 Q(t)+\phi_2 \right)\biggr] \frac{e^{j\Omega t}}{2}\ \nonumber\\
    &\approx\biggr[ \mu_1 I(t)+j\mu_2 Q(t) +\phi_1+j\phi_2 \biggr] \frac{e^{j\Omega t}}{2}\ \nonumber\\
     & \approx \left(\frac{ \mu}{2}\alpha(t) + \Delta\right) e^{j\Omega t}\ ,
    \label{Eq:5}
 \end{align}
 where $\phi_1$ and $\phi_2$ are phase error due to bias deviation from $V_\pi$ of the MZMs. From equation (\ref{Eq:5}), where the approximation $\sin(\theta) \approx \theta$ is justified due to the low RF modulation depth and minor deviations from the expected DC bias values, it is clear that the output of the IQ modulator is side-channel free and the modulated signal is mixed with the optical carrier component $\Delta\exp(j\Omega t)$ due to the finite carrier suppression, as presented in Fig.~\ref{fig:1}(e). This carrier component can be easily removed using a digital filter, as we will explain later. Therefore, the use of baseband modulation for quantum state preparation can remove the sideband leakage vulnerability while keeping the advantages of OSSB in high spectral efficiency and better noise performance by avoiding the low frequency noise using RF heterodyne detection. 
    
\section{Experiment}
\label{sec: Experiment }

 \begin{figure*}[t]
\centering
%\fbox{\includegraphics[width=\linewidth]{IQmodulator.pdf}}
\includegraphics[width=\linewidth]{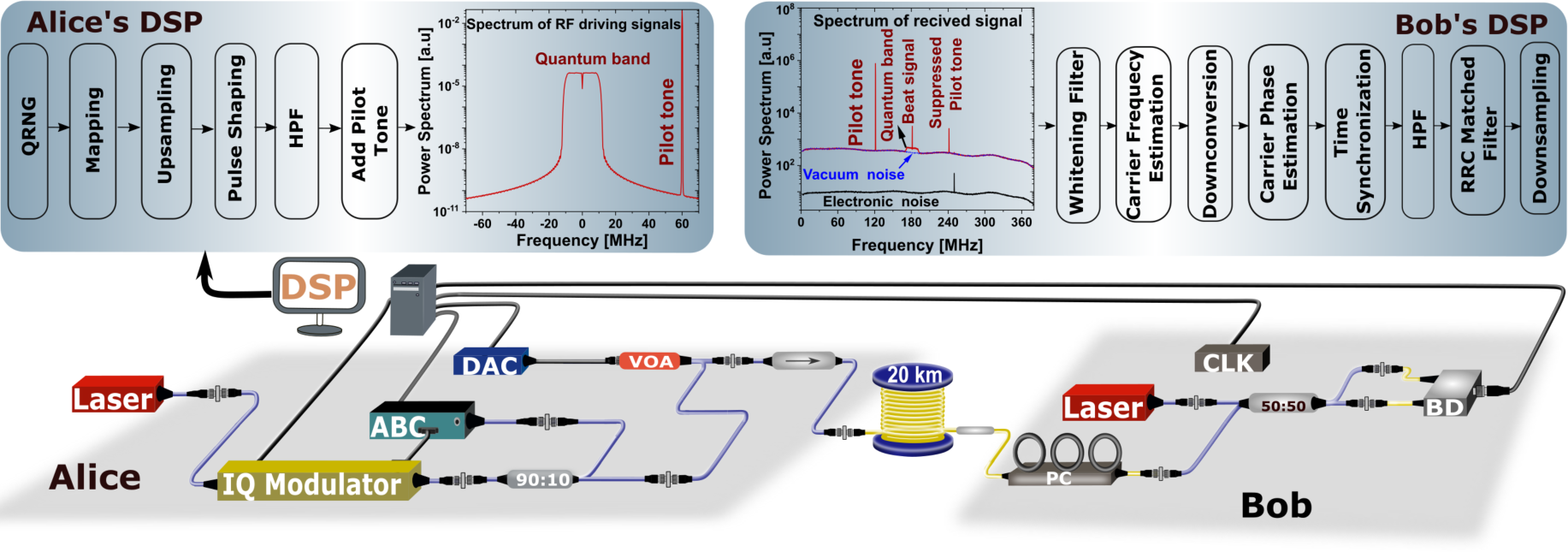}
\caption{Optical layout and DSP routine of our modulation-leakage-free CVQKD system. Alice digitally prepared an ensemble of Gaussian-modulated coherent states with a frequency multiplexed pilot tone using her DSP module. The spectrum of the complex digital signal is shown in the inset. The corresponding RF signals were generated using an arbitrary waveform generator (AWG; not shown in the figure) with 1 GSample/s sampling rate. These signals were then used to drive an IQ modulator modulating light from a 1550 nm continuous-wave laser. For optical carrier suppression, an automatic bias controller (ABC) was used with the IQ modulator. The quantum signal was obtained after the output of IQ modulator was suitably attenuated using a variable optical attenuator (VOA; input and output pigtails on the same side) driven by a digital-to-analog converter (DAC). At Bob's station the polarization of the signal transmitted through a 20 km SMF-28 fiber was corrected with a manual polarization controller (PC) to match the polarization of  an independent local oscillator for RF heterodyne detection. The digitized output of the balanced detector (BD) was fed through Bob's DSP module. The ADC output spectra from 3 different measurements are shown as the input to Bob's DSP routine. An external 10 MHz clock reference (CLK) was used to synchronize the AWG and ADC.} 
\label{fig:2}
\end{figure*}
 
 The schematic of our CVQKD system is shown in Fig.~\ref{fig:2}. It consists of an optical module and a DSP module, each of which are explained below in detail for both the sender and receiver. 
\subsection{Sender (Alice)}
 We digitally generated baseband RF driving signals using Alice's DSP module as shown in Fig.~\ref{fig:2}. As a first step, a quantum random number generator (QRNG) with a security parameter $\epsilon_\text{qrng}= 2 \times 10^{-6}$  was employed to generate uniformly distributed random bit sequences \cite{gehring2021homodyne}. Then, the inversion sampling method based on the cumulative distribution function was used to map  the uniformly distributed sequences  to Gaussian distributed integers with 6 bits resolution covering 7 standard deviations \cite{jain2021practical}. Afterwards, the quantum symbols $A_i = I_i+jQ_i $ were drawn from  these Gaussian distributed integers at rate B = 20 MBaud. These symbols were upsampled to 1 GSamples/s and pulse-shaped by a root raised cosine (RRC) filter with a roll-off factor of 0.2 to obtain the baseband quantum signal $q(t) = h(t)*\sum_i A_i \delta(t-iT) $. Here, $ h(t) $ is the impulse response of the RRC filter. A 5th-order butterworth high pass filter (HPF) with a cut-off frequency of 190 kHz was applied to the quantum signal for the purpose of temporal mode shaping. A reference pilot tone at a frequency of 60 MHz was multiplexed with the quantum signal for frequency offset estimation and phase noise compensation. Figure~\ref{fig:2} (left inset) shows the complex spectrum of $V(t)$, the output of Alice's DSP module. Finally, the RF driving waveforms, $V_\text{RF1}(t) = \text{Real}(V(t))$ and $V_\text{RF2}(t) = \text{Imag}(V(t))$ were uploaded to a dual-channel arbitrary waveform generator (AWG) with 16 bit resolution and sampling frequency of 1 GSample/s. 
 
 In the optical module, a continuous wave (CW) laser at 1550 nm  with a line-width  of $ \approx $ 100 Hz was used as the optical carrier. Random coherent states were prepared using a commercial off-the-shelf IQ modulator driven by the AWG. The DC bias voltages of the IQ modulator were controlled using a commercial ABC to achieve optical baseband modulation with finite carrier suppression. For such an ABC the amount of carrier suppression can be controlled by tuning parameters such as the dither signal amplitude and the feedback photodiode gain of the bias circuit. After the IQ modulator the optical signal was suitably attenuated using an electro-optic variable optical attenuator (VOA) driven by a digital-to-analog converter. To avoid back reflections and thus Trojan-horse attacks \cite{jain2014risk}, a Faraday isolator was added before the input of the quantum channel, which was a 20 km standard single mode fiber.
 
\subsection{Receiver (Bob)}
 After the quantum channel, a manual polarization controller was used to optimize the polarization of the optical signal for RF heterodyne detection with a free-running local oscillator (LO), generated by a CW laser with a frequency offset of $\approx $ 200 MHz with respect to Alice's laser. A home-made broadband balanced detector with a bandwidth of $ \approx$ 365 MHz detected the interference between the LO and the transmitted optical signal. The output of the balanced detector was then digitized at a sampling rate of 1 GSample/s using an analog-to-digital converter (ADC), which was clock synchronised to Alice's AWG using a 10 MHz external reference. 

 The measurement time was divided into frames, each containing $10^7$ samples. Three individual measurements were performed. In order, these were: modulated signal measurement, vacuum noise measurement where Alice's laser was switched off while Bob's laser was on, and finally an electronic noise measurement in which both Alice's and Bob's lasers were switched off. Figure~\ref{fig:2} (right inset) shows the labelled spectral traces from these measurements, evaluated for one frame each, as input to Bob's DSP module. The clearance of the vacuum noise with respect to the electronic noise was $ \approx 15 $ dB. It should also be mentioned that for the modulation strength ($V_a$) calibration, we performed a back-to-back measurement, i.e.\ the transmitter and the receiver were directly connected without the quantum channel. The modulation strength, described as the mean photon number of a thermal state, was $\approx$ 0.27 PNU, where PNU stands for photon number units~\cite{jain2021practical,weedbrook2012gaussian}. 
 
 To recover Alice's modulated quantum symbols, Bob’s DSP module, shown in Fig.~\ref{fig:2}, was applied offline. Whitening filter coefficients were first created by taking the inverse Fourier transform of the averaged frequency response of the vacuum noise. The whitening filter was then applied to electronic, vacuum and quantum signal traces to obtain a flat response across the entire spectrum. The frequency offset was estimated by means of a Hilbert transform of the pilot tone and a linear fit of the extracted phase profile. Using the estimated frequency offset, the pilot tone and the quantum signal were downconverted to baseband. The baseband pilot signal was used as input to an unscented Kalman filter (UKF) for carrier phase estimation~\cite{chin2021machine}. Afterwards, the quantum signal was corrected by the obtained phase estimate from the UKF. Temporal synchronization was achieved through the cross correlation between transmitted and received reference symbols. As a result of baseband modulation, the baseband quantum signal contained undesirable low frequency components, namely, from the optical carrier (beat signal) and the ABC dither signals. To remove these components and measure the same  temporal mode as the transmitted coherent states, a HPF with same type, order and bandwidth as Alice's HPF was applied to the quantum signal as well as the vacuum and electronic noise measurements. We should mention that the bandwidth of the HPFs mainly relies on the spectral width of the beat signal, and the dither frequencies of the ABC. Finally, the received quantum symbols were obtained after RRC matched filtering and downsampling.
 
\subsection{Information reconciliation, parameter estimation and privacy amplification}
After the quantum symbol recovery, Alice and Bob performed information reconciliation (IR) based on a multi-dimensional (MD) scheme using a multi-edge-type low-density-parity-check (MET-LDPC) error correcting code \cite{mani2021multiedge}, assuming they were connected through an authenticated classical channel. Table~\ref{table:code_dd} summarizes the related parameters of the constructed MET-LDPC code, optimized for a binary input additive white Gaussian noise (BI-AWGN) channel. The asymptotic threshold ($\sigma^*_\textsubscript{DE}$) of 5.93 was estimated by running the density evolution until the error probability was $< 10^{-10}$. The code rate (R) of the designed MET-LDPC code was  $0.02$. 

To achieve a high IR efficiency with this code rate at a non-optimal signal-to-noise ratio (SNR) of 0.0443, we adopted the rate-adaptive reconciliation protocol, using a puncturing technique to change the code rate of the designed MET-LDPC code \cite{mani2021multiedge}. By adding punctured symbols whose values between Alice and Bob are uncorrelated, the mutual information  between Alice and Bob data decreases, allowing us to use an error correction code of rate 0.02. The overall MD reconciliation efficiency with puncturing for a dimension $\text{dim}=8$ was computed as $\beta = {R_\text{punc}}/{C_{\text{AWGN}}(s)}$, where $R_\text{punc}$ is the code rate after puncturing. For a codeword of length $n$ and $k$ information bits the original code rate is $R = {k}/{n}$, while after puncturing the final code rate becomes  $R_\text{punc}={k}/{(n-p)}~,$ where $p$ denotes the puncturing length. Table~\ref{tab:puncturing} shows simulated IR efficiencies and frame error rates (FERs) for different puncturing lengths. High IR efficiencies up to 96.46 \% can be achieved at the cost of high FER. Therefore, as a trade-off between the FER and efficiency, we set the puncturing length to 320000, which corresponds to an efficiency of 93.04 \% and FER = 0.215.

% Puncturing technique \cite{PJ-PhysRevA.84.062317} is used here to bring the noise level of the data higher to be able to use an error correction code of rate 0.02. By puncturing some additional noise are then injected between Alice and Bob's data which relatively decrease the SNR between the two parties. The important point is that still the same amount of information can be extracted with an error correction code with lower rate.  

% It is clear that by adding punctured bits whose the value between Alice and Bob are uncorrelated, the SNR between Alice and Bob data increases, which will relatively increase the code rate depending on the puncturing length. 

\begin{table*}[!ht]
\caption{\textbf{Parameters of the designed MET-LDPC code}.   $\sigma^*_\textsubscript{Sh}$ denotes the threshold at Shannon capacity at which the SNR is specified.  $\beta_\text{Code}^*$ is the asymptotic code efficiency.} 
	\begin{center}\label{table:code_dd}
% 	\resizebox{0.98\hsize}{!}{
		\begin{tabular}{cccccc}
			\hline
			$R$ & Degree distribution& $\sigma^*_\textsubscript{DE}$& SNR [dB]&$\sigma^*_\textsubscript{Sh}$&$\beta_\text{Code}^*$\\
                        \hline
            \multirow{2}{*}{0.02}&$\nu(\mathbf{r,x}) =~0.0225~r_1x_1^{2}x_2^{52} + 0.0175~r_1x_1^{3}x_2^{57} +0.96~r_1x_3^1,$&\multirow{2}{*}{$5.93$}&\multirow{2}{*}{$-15.46$}&\multirow{2}{*}{$5.96$}&\multirow{2}{*}{$98.8\%$}\\
            &$\varrho(\mathbf{x}) = ~0.0165~x_1^{4} + 0.0035~x_1^{9}+0.2475~x_2^{3}x_3^1+ 0.7125~x_2^{2}x_3^1~.$&&&&\\
			\hline
		\end{tabular}
% 		}
	\end{center}
\end{table*}

\begin{table}[htp]
\caption{\textbf{The reconciliation efficiency and FER for different puncturing lengths}.The original code has a rate $R= 0.02$, with codeword length $n=1.024\times10^6$ and information length $k = 20480$.}\begin{center}
    \label{tab:puncturing}
    \begin{tabular}{c|c|c|c}
    \hline
         $\beta~(\%)$& FER & $p$ & $R_\text{punc}$\\
         \hline
         92.77& 0.075 & 318000 & 0.0290\\
         \textbf{93.04}& \textbf{0.215} & \textbf{320000} & \textbf{0.0291}\\
         93.70& 0.378 & 325000 & 0.0293\\
         94.37& 0.480 & 330000 & 0.0295\\
         95.06& 0.716 & 335000 & 0.0297\\
         95.75& 0.850 & 340000 & 0.0299\\
         96.46& 0.962 & 345000 & 0.0302
    \end{tabular}
    \end{center}
\end{table}
%\begin{figure}[!ht]
%	\centering
%	\includegraphics[width = 0.45\textwidth, keepaspectratio]{MDwS.pdf}
%	\caption{\textcolor{red}{Optional you may want to remove this figure. } \textbf{Block diagram for IR.} Reverse reconciliation using multidimensional reconciliation scheme. $\vec{x}_A$ and $\vec{x}_B$ are two correlated sequences of real valued data. $\vec{x'}_A$ and $\vec{x'}_B$ denote the corresponding normalized sequences. $\vec{\alpha} = M(\vec{x'}_B, \vec{c})$ represents the mapping function. The raw key is generated by a QRNG is denoted by $\vec{c}$. The non-zero syndrome of the key calculated using the parity check of the MET-LDPC code and transmitted through the authenticated channel is denoted by $\vec{s}$.}\label{fig:md-recon-non-zero-synd}
% \end{figure}

Next, Alice performed parameter estimation to bound Eve’s Holevo information, calculating the number of bits expected in the output secret key in the worst-case scenario. This length was communicated together with a seed to Bob. For privacy amplification, the shared seed from the previous step was used to select a random Toeplitz hash function by Alice and Bob, who then employed the high-speed and large-scale PA scheme \cite{Tang2019} to generate the final secret key.

% ~\cite{milicevic2018quasi}, 
% $%\begin{equation}
%     \beta = \beta_\text{Code} \times \beta_\text{Channel} = \beta_\text{Code} \times   \dfrac{C_{\text{dim}=8}(s)}{C_{\text{AWGN}}(s)}$,
% %\end{equation}
% where $\beta_\text{Code}$ is the code efficiency in practice (due to the finite length used in the error correction). At the estimated SNR = 0.193, we obtained $\beta_\text{Code} = 94.31$ \% and $\beta_\text{Channel} = 99.9$ \%, yielding an overall reconciliation efficiency of $94.26$ \%. After IR, Alice performed parameter estimation to bound Eve’s Holevo information. Finally, both parties performed privacy amplification.
% %############################################%
% %############################################%

 %\vspace{-1.5cm}
\section{Results}
\begin{figure}[!ht]
\centering
%\fbox{\includegraphics[width=\linewidth]{IQmodulator.pdf}}
\includegraphics[width=\linewidth]{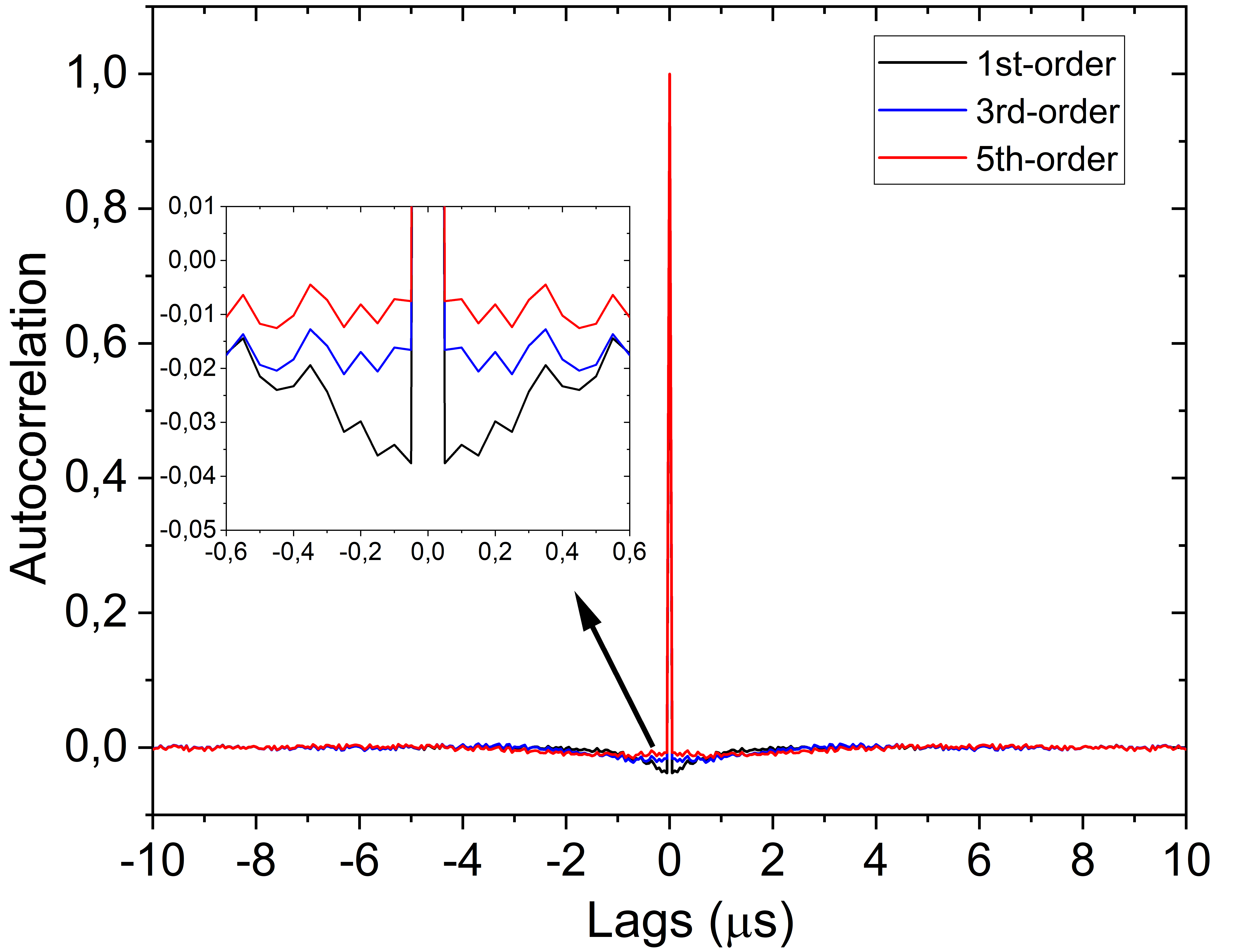}
\caption{ Normalized autocorrelation function of demodulated quantum symbols using HPFs of the same bandwidth and different orders.}
\label{fig:3}
\end{figure}

The use of HPFs in Alice's and Bob's DSP modules is the key ingredient of our CVQKD implementation. To understand the effect of these HPFs on the security and system performance, we first investigated the normalized auto-correlation functions of received quantum symbols from one frame, demodulated using three different HPFs with the same bandwidth of 190 kHz but different orders, as depicted in Fig.~\ref{fig:3}. In general, applying these HPFs can introduce correlations between consecutive symbols as indicated by the dip around the correlation peak, clearly seen in the figure in the case of the 1st-order filter. With reference to the spectrum of the RF driving signal in Fig.~\ref{fig:2}, this correlation could be attributed to the fact that the HPF gives rise to frequency-selective fading, indicated by the notch in the spectrum, and causes inter‐symbol interference in the time domain \cite{armstrong2009ofdm}. Moreover, the inset in Fig.~\ref{fig:3} shows that the strength of correlation depends on the filter order: as the order reduces the dip in correlation increases. Therefore, one could also anticipate the same dependence for the filter bandwidth since the delay spread is proportional to the  notch bandwidth  in  the spectrum introduced by the HPF \cite{armstrong2009ofdm}. 

From a security point of view, such correlations can destroy the independent and identically distributed property of the quantum symbols, which would violate an assumption commonly made in the security proofs. Therefore, optimizing filter parameters is extremely important to keep the effect on the independence as small as possible. However, we note, that even without HPF perfect independence cannot be achieved in practice due to the physical properties of the modulator. One way the results suggest for filter optimization is to minimize the correlation coefficient in the vicinity of the correlation peak, as in the case of the 5th-order filter.
 
\begin{figure}[!h]
\centering
%\fbox{\includegraphics[width=\linewidth]{IQmodulator.pdf}}
\includegraphics[width=\linewidth]{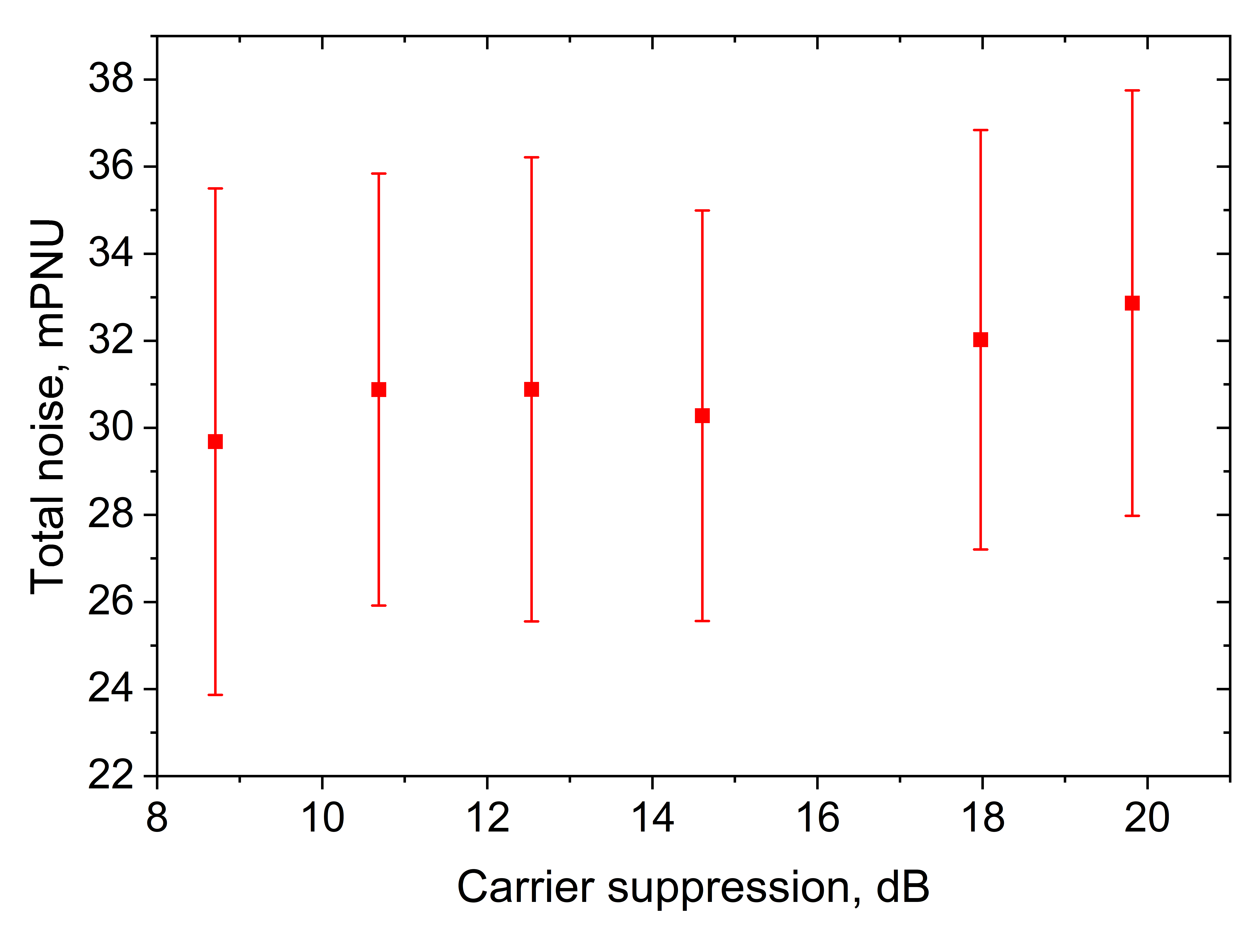}
\caption{Total noise versus carrier suppression. }
\label{fig:4}
\end{figure}

 \begin{figure*}[!t]
\centering
%\fbox{\includegraphics[width=\linewidth]{IQmodulator.pdf}}
\includegraphics[scale=0.42]{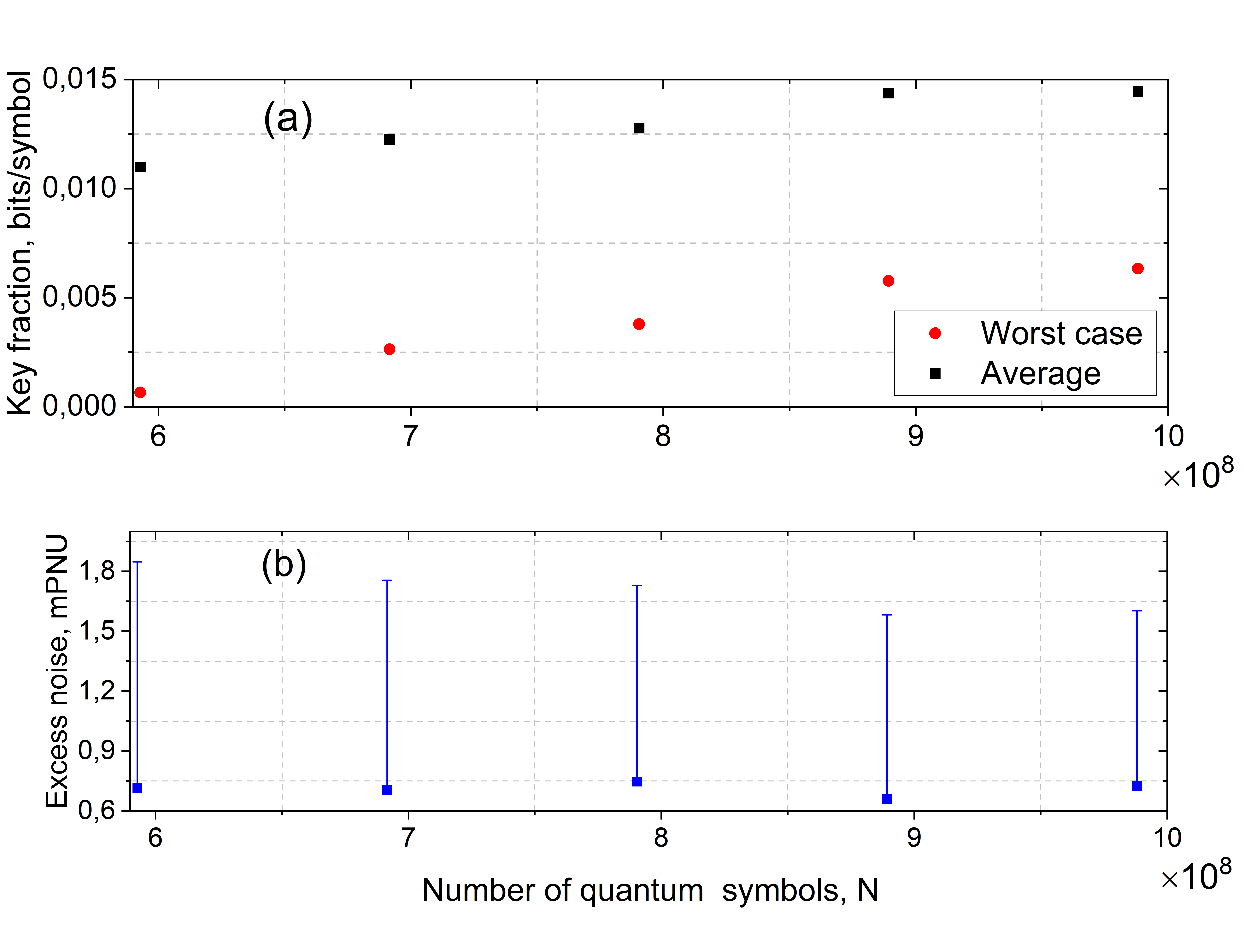}
\caption{Experimental results demonstrating the CVQKD system performance. (a) Composable secret key fraction versus number of  quantum symbols. (b) Measured excess noise and its worst case estimator.}
\label{fig:5}
\end{figure*} 
As a next step, we explored the tolerance of the optimized HPFs to finite carrier suppression provided by the ABC. To do this, we performed regular heterodyne measurements while varying the amount of carrier suppression by tuning the dither amplitude applied by the ABC on MZM1 and MZM2. As a figure of merit for the HPF's performance, Fig.~\ref{fig:4} shows the total noise, i.e.\ the sum of the trusted electronic noise ($t$) and the untrusted noise ($u$) contributed by Eve, as function of the carrier suppression. Each data-point in the plot is represented by the mean and standard deviation over 200 frames. It is clear that the HPF has no significant performance degradation over 10 dB carrier suppression, implied by the highly overlapping error bars. This result confirms the effectiveness of the optimized HPF over a relatively wide range of carrier suppression. We note that as the ABC cannot provide lower than 9 dB carrier suppression, we were not able to reach the performance breaking point.

Finally, to demonstrate the practical feasibility of modulation-leakage-free CVQKD, we generated a secret key over the 20\,km fiber considering composable security and collective attacks according to \cite{jain2021practical}. Table \ref{tab:1} summarizes the main experimental parameters used for secret key generation. Alice sent $10^9$ coherent states with a modulation strength of 270 mPNU to Bob. After channel parameter estimation, a total noise $u+t = 0.73+31.40 = 32.13$ mPNU and a total average transmittance  $\eta \cdot\tau = 0.24 \cdot0.68 = 0.16$ were measured. Here, $\eta$ is the untrusted channel transmission and $\tau$ is the trusted detection efficiency. In comparison to \cite{jain2021practical} this low excess noise was achieved because our system operated at a lower modulation strength, where phase noise is not a dominant factor. Also, careful optimization of our DSP chain played a significant role to obtain such noise performance. We note that $\eta$ is  relatively small for a 20 km fiber channel with a physical transmission of $\approx 0.38$, which could be attributed to the polarization optimization after the fiber channel.

As for IR, Alice and Bob used $987648000 $ complex symbols to perform reverse reconciliation with an efficiency  $ \beta = 93.04 \% $ with the IR code described above in frames of $1024000$ bits. The total number of corrected symbols after IR was reduced to $ 775303680 $ due to FER = $ 0.215$. The maximum iteration number for decoding of the MET-LDPC decoder was set to 500. Joining data from both I and Q quadratures, we obtained a total of $2\times775303680 $ symbols. Finally, we calculated the composable secret key length $l = 10854251$ bits, which translates into a secret key fraction of $0.007$ bits/symbol. The security parameters of calibration ($\epsilon_\text{cal}$) and parameter estimation ($\epsilon_\text{PE}$) were both set to $ 10^{-10} $ while $\epsilon_{IR}$ was set to $ 10^{-12}$. 

\begin{table}[t]
\centering
\caption{\textbf{ Overview of experimental parameters for composable security}. $\tau$: Trusted efficiency, $\eta$: Untrusted efficiency, $t$: trusted detection noise, $u$: untrusted channel noise, FER: frame error rate, $\beta$: IR efficiency.}
\resizebox{0.98\hsize}{!}{
\begin{tabular}{cccc}
\hline
\bf Alice & \bf Bob&\bf channel &\bf IR\\
\hline
$B = 20$ MBaud &$\tau=0.68$ & $\eta=0.24$ & FER $=0.215$\\
$V_a = 0.27$ PNU &$t=31.40$ mPNU &$u=0.73$ mPNU &$\beta = 93.04\%$ \\
\hline
\end{tabular}
}
  \label{tab:1}
\end{table}

Fig.~\ref{fig:5} (a) shows the composable secret key fraction as a function of the number $N$ of available quantum symbols (before IR) both for the average (black squares) and worst-case (red dots) scenarios -- the latter given by the considered confidence intervals~\cite{jain2021practical}. Likewise, the measured untrusted noise with worst case estimator is shown in Fig.~\ref{fig:5} (b). %Note that the x-axes of Fig.~\ref{fig:5} can also be considered as a time evolution with interval $ T = 1/B$ between the symbols, if the time taken by data acquisition, DSP and the protocol steps are neglected.
A positive key fraction was obtained for $N \gtrsim 6\times 10^8 $, corresponding to a null key fraction threshold of $\approx 1.8$ mPNU. From an experimental point of view, the reason for being able to achieve a positive key fraction with a relatively  small number of symbols could be attributed to our system noise performance, implying the practical feasible of our implementation. Also considering an estimated $\eta$ of 0.24, a composable secret key for a longer distance can be achieved by optimizing the polarization control circuit.     

\section{DISCUSSION}
 In this work we demonstrated a practical CVQKD system which is  modulation leakage free. This was enabled by means of optical baseband modulation for coherent states preparation and RF heterodyne reception together with carefully designed DSP. Our system showed low noise performance with high tolerance to finite carrier suppression of the IQ modulator due to the use of digital HPFs. This allowed us to generate a composable secret key secure against collective attacks over 20 km fiber. However, a careful optimization of the HPFs is necessary to avoid inter symbol interference, that may leak more information to Eve. 
 
 Compared with phase-diverse intradyne or homodyne CVQKD systems, where the security is not affected by the sideband modulation leakage as Bob measures both sidebands, our system provides a cost-effective solution since it does not require a 90 degree optical hybrid nor an amplitude modulator in the transmitter to generate pulses. Besides, like OSSB CVQKD systems, our implementation facilitates a spectral efficiency and better noise performance by avoiding the low frequency noise at detection side. The proposed baseband modulation together with the DSP chain is also suitable for other CVQKD systems based on modulated coherent states, including point-to-point and measurement-device-independent CVQKD. %Overall, with regards to the considerable effort that has been devoted by public institutions, standardization bodies and companies \cite{marco2018implementation} to practical security of QKD, 
 We anticipate that our implementation can be a potential solution to protect future CVQKD systems against side-channels from modulation leakage with better system performance, and remove the need for a complex characterization of the leakage and the risk of a failure in system security due to long-term stability issues which may increase the leakage over time.

\begin{backmatter}
\vspace{0.5cm}
\bmsection{Acknowledgments} The authors acknowledge support from Innovation Fund Denmark (CryptQ, grant agreement no. 0175-00018A), from European Union’s Horizon 2020 research and innovation programmes Uniqorn (grant agreement no.\ 820474), CiViQ (grant agreement no. 820466) and OPENQKD (grant agreement no.\ 857156), from the Independent Research Fund Denmark (grant agreement no.\ 0171-00055B) and from the Danish National Research Foundation, Center for Macroscopic Quantum States (bigQ, DNRF142).

\smallskip

\bmsection{Disclosures} The authors declare no conflicts of interest.

\smallskip

\bmsection{Data availability} Data underlying the results presented in this paper are available on request.

\end{backmatter}

\bigskip

% Bibliography
\bibliography{sample}

\begin{thebibliography}{10}
\newcommand{\enquote}[1]{``#1''}

\bibitem{scarani2009security}
V.~Scarani, H.~Bechmann-Pasquinucci, N.~J. Cerf, M.~Du{\v{s}}ek,
  N.~L{\"u}tkenhaus, and M.~Peev, \enquote{The security of practical quantum
  key distribution,} {\protect\JournalTitle{Reviews of Modern Physics}}
  \textbf{81}, 1301 (2009).

\bibitem{pirandola2020advances}
S.~Pirandola, U.~L. Andersen, L.~Banchi, M.~Berta, D.~Bunandar, R.~Colbeck,
  D.~Englund, T.~Gehring, C.~Lupo, C.~Ottaviani \emph{et~al.},
  \enquote{Advances in quantum cryptography,} {\protect\JournalTitle{Advances
  in Optics and Photonics}} \textbf{12}, 1012--1236 (2020).

\bibitem{Gisin2006}
N.~Gisin, S.~Fasel, B.~Kraus, H.~Zbinden, and G.~Ribordy,
  \enquote{{Trojan-horse attacks on quantum-key-distribution systems},}
  {\protect\JournalTitle{Physical Review A}} \textbf{73}, 022320 (2006).

\bibitem{jain2011device}
N.~Jain, C.~Wittmann, L.~Lydersen, C.~Wiechers, D.~Elser, C.~Marquardt,
  V.~Makarov, and G.~Leuchs, \enquote{Device calibration impacts security of
  quantum key distribution,} {\protect\JournalTitle{Physical Review Letters}}
  \textbf{107}, 110501 (2011).

\bibitem{lydersen2010hacking}
L.~Lydersen, C.~Wiechers, C.~Wittmann, D.~Elser, J.~Skaar, and V.~Makarov,
  \enquote{Hacking commercial quantum cryptography systems by tailored bright
  illumination,} {\protect\JournalTitle{Nature Photonics}} \textbf{4}, 686--689
  (2010).

\bibitem{gerhardt2011full}
I.~Gerhardt, Q.~Liu, A.~Lamas-Linares, J.~Skaar, C.~Kurtsiefer, and V.~Makarov,
  \enquote{Full-field implementation of a perfect eavesdropper on a quantum
  cryptography system,} {\protect\JournalTitle{Nature Communications}}
  \textbf{2}, 1--6 (2011).

\bibitem{zhang2020long}
Y.~Zhang, Z.~Chen, S.~Pirandola, X.~Wang, C.~Zhou, B.~Chu, Y.~Zhao, B.~Xu,
  S.~Yu, and H.~Guo, \enquote{Long-distance continuous-variable quantum key
  distribution over 202.81 km of fiber,} {\protect\JournalTitle{Physical Review
  Letters}} \textbf{125}, 010502 (2020).

\bibitem{grosshans2002continuous}
F.~Grosshans and P.~Grangier, \enquote{Continuous variable quantum cryptography
  using coherent states,} {\protect\JournalTitle{Physical Review Letters}}
  \textbf{88}, 057902 (2002).

\bibitem{weedbrook2004quantum}
C.~Weedbrook, A.~M. Lance, W.~P. Bowen, T.~Symul, T.~C. Ralph, and P.~K. Lam,
  \enquote{Quantum cryptography without switching,}
  {\protect\JournalTitle{Physical Review Letters}} \textbf{93}, 170504 (2004).

\bibitem{laudenbach2018continuous}
F.~Laudenbach, C.~Pacher, C.-H.~F. Fung, A.~Poppe, M.~Peev, B.~Schrenk,
  M.~Hentschel, P.~Walther, and H.~H{\"u}bel, \enquote{Continuous-variable
  quantum key distribution with gaussian modulation—the theory of practical
  implementations,} {\protect\JournalTitle{Advanced Quantum Technologies}}
  \textbf{1}, 1800011 (2018).

\bibitem{jouguet2013experimental}
P.~Jouguet, S.~Kunz-Jacques, A.~Leverrier, P.~Grangier, and E.~Diamanti,
  \enquote{Experimental demonstration of long-distance continuous-variable
  quantum key distribution,} {\protect\JournalTitle{Nature Photonics}}
  \textbf{7}, 378--381 (2013).

\bibitem{weedbrook2012gaussian}
C.~Weedbrook, S.~Pirandola, R.~Garc{\'\i}a-Patr{\'o}n, N.~J. Cerf, T.~C. Ralph,
  J.~H. Shapiro, and S.~Lloyd, \enquote{Gaussian quantum information,}
  {\protect\JournalTitle{Reviews of Modern Physics}} \textbf{84}, 621 (2012).

\bibitem{kikuchi2015fundamentals}
K.~Kikuchi, \enquote{Fundamentals of coherent optical fiber communications,}
  {\protect\JournalTitle{Journal of Lightwave Technology}} \textbf{34},
  157--179 (2015).

\bibitem{ma2013local}
X.-C. Ma, S.-H. Sun, M.-S. Jiang, and L.-M. Liang, \enquote{Local oscillator
  fluctuation opens a loophole for eve in practical continuous-variable
  quantum-key-distribution systems,} {\protect\JournalTitle{Physical Review A}}
  \textbf{88}, 022339 (2013).

\bibitem{jouguet2013preventing}
P.~Jouguet, S.~Kunz-Jacques, and E.~Diamanti, \enquote{Preventing calibration
  attacks on the local oscillator in continuous-variable quantum key
  distribution,} {\protect\JournalTitle{Physical Review A}} \textbf{87}, 062313
  (2013).

\bibitem{qin2016quantum}
H.~Qin, R.~Kumar, and R.~All{\'e}aume, \enquote{Quantum hacking: Saturation
  attack on practical continuous-variable quantum key distribution,}
  {\protect\JournalTitle{Physical Review A}} \textbf{94}, 012325 (2016).

\bibitem{huang2013quantum}
J.-Z. Huang, C.~Weedbrook, Z.-Q. Yin, S.~Wang, H.-W. Li, W.~Chen, G.-C. Guo,
  and Z.-F. Han, \enquote{Quantum hacking of a continuous-variable
  quantum-key-distribution system using a wavelength attack,}
  {\protect\JournalTitle{Physical Review A}} \textbf{87}, 062329 (2013).

\bibitem{derkach2017continuous}
I.~Derkach, V.~C. Usenko, and R.~Filip, \enquote{Continuous-variable quantum
  key distribution with a leakage from state preparation,}
  {\protect\JournalTitle{Physical Review A}} \textbf{96}, 062309 (2017).

\bibitem{jain2021modulation}
N.~Jain, I.~Derkach, H.-M. Chin, R.~Filip, U.~L. Andersen, V.~C. Usenko, and
  T.~Gehring, \enquote{Modulation leakage vulnerability in continuous-variable
  quantum key distribution,} {\protect\JournalTitle{Quantum Science and
  Technology}} \textbf{6}, 045001 (2021).

\bibitem{kleis2017continuous}
S.~Kleis, M.~Rueckmann, and C.~G. Schaeffer, \enquote{Continuous variable
  quantum key distribution with a real local oscillator using simultaneous
  pilot signals,} {\protect\JournalTitle{Optics Letters}} \textbf{42},
  1588--1591 (2017).

\bibitem{qu2016rf}
Z.~Qu, I.~B. Djordjevic, and M.~A. Neifeld,
  \enquote{\uppercase{RF}-subcarrier-assisted four-state continuous-variable
  qkd based on coherent detection,} {\protect\JournalTitle{Optics Letters}}
  \textbf{41}, 5507--5510 (2016).

\bibitem{chin2021machine}
H.-M. Chin, N.~Jain, D.~Zibar, U.~L. Andersen, and T.~Gehring, \enquote{Machine
  learning aided carrier recovery in continuous-variable quantum key
  distribution,} {\protect\JournalTitle{npj Quantum Information}} \textbf{7},
  1--6 (2021).

\bibitem{brunner2017low}
H.~H. Brunner, L.~C. Comandar, F.~Karinou, S.~Bettelli, D.~Hillerkuss, F.~Fung,
  D.~Wang, S.~Mikroulis, Q.~Yi, M.~Kuschnerov \emph{et~al.}, \enquote{A
  low-complexity heterodyne \uppercase{CV-QKD} architecture,} in \emph{2017
  19th International Conference on Transparent Optical Networks (ICTON),}
  (IEEE, 2017), pp. 1--4.

\bibitem{gehring2015sqzcomposableqkd}
T.~Gehring, V.~H{\"{a}}ndchen, J.~Duhme, F.~Furrer, T.~Franz, C.~Pacher, R.~F.
  Werner, and R.~Schnabel, \enquote{{Implementation of continuous-variable
  quantum key distribution with composable and one-sided-device-independent
  security against coherent attacks},} {\protect\JournalTitle{Nature
  Communications}} \textbf{6}, 1--7 (2015).

\bibitem{jain2021practical}
N.~Jain, H.-M. Chin, H.~Mani, C.~Lupo, D.~S. Nikolic, A.~Kordts, S.~Pirandola,
  T.~B. Pedersen, M.~Kolb, B.~{\"O}mer \emph{et~al.}, \enquote{Practical
  continuous-variable quantum key distribution with composable security,}
  {\protect\JournalTitle{arXiv:2110.09262}}  (2021).

\bibitem{li2014continuous}
Z.~Li, Y.-C. Zhang, F.~Xu, X.~Peng, and H.~Guo, \enquote{Continuous-variable
  measurement-device-independent quantum key distribution,}
  {\protect\JournalTitle{Physical Review A}} \textbf{89}, 052301 (2014).

\bibitem{lance2005no}
A.~M. Lance, T.~Symul, V.~Sharma, C.~Weedbrook, T.~C. Ralph, and P.~K. Lam,
  \enquote{No-switching quantum key distribution using broadband modulated
  coherent light,} {\protect\JournalTitle{Physical Review Letters}}
  \textbf{95}, 180503 (2005).

\bibitem{jain2014risk}
N.~Jain, B.~Stiller, I.~Khan, V.~Makarov, C.~Marquardt, and G.~Leuchs,
  \enquote{Risk analysis of \uppercase{T}rojan-horse attacks on practical
  quantum key distribution systems,} {\protect\JournalTitle{IEEE Journal of
  Selected Topics in Quantum Electronics}} \textbf{21}, 168--177 (2014).

\bibitem{Izutsu1981IntegratedIQmod}
M.~Izutsu, S.~Shikama, and T.~Sueta, \enquote{Integrated optical ssb
  modulator/frequency shifter,} {\protect\JournalTitle{IEEE Journal of Quantum
  Electronics}} \textbf{17}, 2225--2227 (1981).

\bibitem{li2017modulation}
X.~Li, L.~Deng, X.~Chen, M.~Cheng, S.~Fu, M.~Tang, and D.~Liu,
  \enquote{Modulation-format-free and automatic bias control for optical
  \uppercase{IQ} modulators based on dither-correlation detection,}
  {\protect\JournalTitle{Optics Express}} \textbf{25}, 9333--9345 (2017).

\bibitem{agrawal2012fiber}
G.~P. Agrawal, \emph{Fiber-optic communication systems} (John Wiley \& Sons,
  2012).

\bibitem{gehring2021homodyne}
T.~Gehring, C.~Lupo, A.~Kordts, D.~S. Nikolic, N.~Jain, T.~Rydberg, T.~B.
  Pedersen, S.~Pirandola, and U.~L. Andersen, \enquote{Homodyne-based quantum
  random number generator at 2.9 \uppercase{G}bps secure against quantum
  side-information,} {\protect\JournalTitle{Nature Communications}}
  \textbf{12}, 1--11 (2021).

\bibitem{mani2021multiedge}
H.~Mani, T.~Gehring, P.~Grabenweger, B.~{\"O}mer, C.~Pacher, and U.~L.
  Andersen, \enquote{Multiedge-type low-density parity-check codes for
  continuous-variable quantum key distribution,}
  {\protect\JournalTitle{Physical Review A}} \textbf{103}, 062419 (2021).

\bibitem{Tang2019}
B.-Y. Tang, B.~Liu, Y.-P. Zhai, C.-Q. Wu, and W.-R. Yu, \enquote{High-speed and
  large-scale privacy amplification scheme for quantum key distribution,}
  {\protect\JournalTitle{Scientific Reports}} \textbf{9}, 15733 (2019).

\bibitem{armstrong2009ofdm}
J.~Armstrong, \enquote{\uppercase{OFDM} for optical communications,}
  {\protect\JournalTitle{Journal of Lightwave Technology}} \textbf{27},
  189--204 (2009).

\end{thebibliography}

% Full bibliography added automatically for Optics Letters submissions; the following line will simply be ignored if submitting to other journals.
% Note that this extra page will not count against page length
\bibliographyfullrefs{sample}

%Manual citation list
%\begin{thebibliography}{1}
%\bibitem{Zhang:14}
%Y.~Zhang, S.~Qiao, L.~Sun, Q.~W. Shi, W.~Huang, %L.~Li, and Z.~Yang,
 % \enquote{Photoinduced active terahertz metamaterials with nanostructured
  %vanadium dioxide film deposited by sol-gel method,} Opt. Express \textbf{22},
  %11070--11078 (2014).
%\end{thebibliography}

% Please include bios and photos of all authors for aop articles
\ifthenelse{\equal{\journalref}{aop}}{%
\section*{Author Biographies}
\begingroup
\setlength\intextsep{0pt}
\begin{minipage}[t][6.3cm][t]{1.0\textwidth} % Adjust height [6.3cm] as required for separation of bio photos.
  \begin{wrapfigure}{L}{0.25\textwidth}
    \includegraphics[width=0.25\textwidth]{john_smith.eps}
  \end{wrapfigure}
  \noindent
  {\bfseries John Smith} received his BSc (Mathematics) in 2000 from The University of Maryland. His research interests include lasers and optics.
\end{minipage}
\begin{minipage}{1.0\textwidth}
  \begin{wrapfigure}{L}{0.25\textwidth}
    \includegraphics[width=0.25\textwidth]{alice_smith.eps}
  \end{wrapfigure}
  \noindent
  {\bfseries Alice Smith} also received her BSc (Mathematics) in 2000 from The University of Maryland. Her research interests also include lasers and optics.
\end{minipage}
\endgroup
}{}

\end{document}